\documentstyle[twocolumn,aps,psfig]{revtex}

\begin{document}
\draft

\twocolumn[\hsize\textwidth\columnwidth\hsize\csname
@twocolumnfalse\endcsname

\title{
Pseudogap Induced Antiferromagnetic Spin Correlation \\in 
High-Temperature Superconductors
}

\author{T\^oru Sakai and Yoshinori Takahashi}
\address{
Faculty of Science, Himeji Institute of Technology, Kamigori,
Ako-gun, Hyogo 678-1297, Japan
}

\date{October 2000}
\maketitle

\begin{abstract}
The pseudogap phenomena observed on cuprate high temperature
  superconductors are investigated based on the exact diagonalization
  method on the finite cluster $t$-$J$ model. The results show the
  presence of the gap-like behavior in the temperature dependence of
  various magnetic properties; the NMR relaxation rate, the neutron
  scattering intensity and the static susceptibility.  The calculated
  spin correlation function indicates that the pseudogap behavior
  arises associated with the development of the antiferromagnetic
  spin correlation with decreasing the temperature.  The numerical
  results are presented to clarify the model parameter dependence,
  that covers the realistic experimental situation. The effect of the
  next-nearest neighbor hopping $t'$ is also studied.
\end{abstract}

\vskip2pc]
\narrowtext

\section{Introduction}

The high-temperature superconductors exhibit various anomalous
metallic properties in their normal phases.  A gap-like behavior,
first discovered by the NMR relaxation rate measurement\cite{nmr1}  as
one of these properties has attracted a lot of current interest. It is
characteristic to the underdoped cuprates and observed as a broad peak
in the $1/T_1T$-$T$ curve at a slightly higher temperature than the
superconducting transition point $T_{\rm c}$.  The possible relation of the
pseudogap phenomena with the origin of the superconductivity underlies
the reason of such interest.  The behavior has now been detected in the
temperature dependence of various other physical quantities; the
neutron scattering intensity\cite{neutron1,neutron2}, the magnetic
susceptibility\cite{chi}, the resistivity\cite{res}, the Hall
coefficient\cite{hall} and the angle-resolved photo-emission spectrum
(ARPES)\cite{arpes}.

A lot of theories have been proposed to explain the phenomena.  Since
they were observed only for compounds with double CuO$_2$ layers like
YBa$_2$Cu$_4$O$_8$ for the first time, the interlayer coupling was
supposed to play the predominant role on  the gap
formation.\cite{bi1,bi2,bi3} However, after the discovery of the
phenomena on the mono-layer oxide HgBa$_2$CuO$_4$ by the NMR
measurement\cite{nmr3}, most theories are based on the model with a
single CuO$_2$ plane, such as the square-lattice $t$-$J$ model which
has been extensively studied theoretically in the description of 
the cuprate superconductivity.\cite{dagotto} Based on the
concept of the spin-charge separation,\cite{anderson} 
Tanamoto {\it et al}.
\cite{tanamoto} showed that the $t$-$J$ model exhibits a gap-like
behavior in its spin excitation spectrum as the result of the spinon
condensation by applying the mean field approximation to the
resonating valence bond basis.  Onoda {\it et al}.\cite{onoda} also derived
the occurrence of the gap-like behavior in the temperature dependence
of the resistivity within the same model using the slave-boson mean
field approximation.  These results, however, crucially depend on the
nature of the approximation, the validity of which is not so well
established.

Miyake and Narikiyo, on the other hand, suggested that the pseudogap
rather originates from the enhancement of the antiferromagnetic spin
fluctuation paying attention on the nesting effect based on the
phenomenological itinerant-localized duality theory.\cite{miyake} In
our previous study, we also proposed a simple explanation based on
the well-known property of the Heisenberg magnets.\cite{sakai} We were 
successful in  demonstrating the presence of the gap-like behavior in
the temperature dependence of the NMR relaxation rate by numerically
diagonalizing the finite cluster $t$-$J$ model for sufficiently large
$J$. Our   result shows it is interpreted as a crossover from the
paramagnetic to the magnetically well correlated state induced by the
rapid development of the antiferromagnetic spin correlation.

Lately lots of interest have been paid on the role of the
superconducting fluctuations.  For instance, Koikegami and
Yamada\cite{koikegami}, based on the $d$-$p$ model, indicated that
superconducting fluctuations will survive at higher temperature than
the transition temperature $T_{\rm c}$ based on the fluctuation exchange
approximation.  This means that the gap-like behavior is regarded as a
precursor of the gap formation of the strong coupling
superconductivity as was predicted by the self-consistent $t$-matrix
approximation by Yanase and Yamada\cite{yanase}.  With the use of the
$t$-matrix approximation Kobayashi {\it et al}\cite{kobayashi} showed that
the gap-like behavior is induced by the appearance of the
superconducting fluctuations while the antiferromagnetic spin
fluctuation is suppressed on the contrary.  Onoda and
Imada\cite{imada} claim that the competition between the
superconducting fluctuations and antiferromagnetic spin fluctuations
is necessary for deriving the gap behavior.

In spite of lots of theoretical proposals, the final answer is still
far from being established. Most theories, however, seem to have an
agreement on the important role of aniferromagnetic spin correlation
on the pseudogap formation.  In the present paper, in order to confirm
the validity of our mechanism we extend our numerical investigation
and derive the temperature dependence of various physical
properties, including $1/T_1T$, in a wider parameter region that
covers realistic situations for high-$T_{\rm c}$ cuprates. The effect of the
next-nearest neighbor hole hopping with a coupling $t'$ is also
studied.  The effect of the antiferromagnetic spin correlation on the
pseudogap formation is then clarified without resorting to any
approximation. If we take into account the lack of reliable
approximation scheme in treating the strongly correlated systems, the
numerical analysis of the model has its own significance in order to
check the validity of various explanations.

\section{Mechanism of Pseudogap}

According to our proposal, the local spin excitations are suppressed
around a characteristic temperature, of the the same magnitude of the
antiferromagnetic coupling $J$, associated with the growth of the
short-range antiferromagnetic correlation as we decrease the
temperature.  As a result various magnetic quantities will show their
particular $T$-dependence in reflecting the reduced local excitations.
In fact the high temperature series expansion for the square-lattice
antiferromagnetic Heisenberg model shows a broad peak in the
temperature dependence of the NMR relaxation rate $1/T_1T$ around
$T\sim J$.\cite{chakravarty} Since the hole motion acts to destroy the
short-range antiferromagnetic order, the hole doping will reduce the
crossover temperature in agreement with the observed hole doping
dependence of the pseudogap temperature.  Thus the pseudogap behavior
is simply realized as the manifestation of the growth of the
short-range antiferromagnetic correlation.  The above picture has been
supported by our numerical exact diagonalization study on the
two-dimensional $t$-$J$ model with 10 lattice sites.\cite{sakai} We
found that for a small hole concentration ($\delta =0.1$) the $T$
dependence of $1/T_1T$ does show a broad peak around the temperature
where the nearest neighbor antiferromagnetic spin correlation shows
rapid growing, for a sufficiently large antiferromagnetic coupling
($J/t=0.6$).  In the next section we perform the same calculation to
investigate how our previous results are modified by  using realistic
parameters for the cuprate superconductors and also to examine the
presence of possible gap-like behaviors for various  magnetic
properties.

\section{Numerical Analysis of $t$-$J$ model}

Let us first consider the standard square-lattice  $t$-$J$ model
Hamiltonian defined by
\begin{eqnarray}
\label{ham}
H &= - t \sum_{<\bf i,\bf j>, \sigma}
            ({c}_{\bf j,\sigma}^\dagger {c}_{\bf i,\sigma}
           + {c}_{\bf i,\sigma}^\dagger {c}_{\bf j,\sigma} ) \nonumber \\
  &+ J \sum_{<\bf i,\bf j>}
           ( {\bf S}_{\bf i} \cdot {\bf S}_{\bf j}
           - \textstyle{1 \over 4} n_{\bf i} n_{\bf j} ) ,
\end{eqnarray}
where $t$ is the nearest neighbor  electron hopping integral and $J$
is the antiferromagnetic Heisenberg exchange constant between spins on
adjacent lattice sites. Throughout the paper, all the energies are
measured in units of $t$. With the use of the calculated eigenvalues
and eigenvectors, we evaluated the temperature  dependence of the
imaginary part of the dynamical spin susceptibility of  conduction
electrons ${\rm Im}\chi (q,\omega)$ and  the static spin correlation
function  $\langle S^z_{\bf i} S^z_{\bf j} \rangle$.  In actual
numerical estimations of ${\rm Im}\chi (q,\omega)$ the
$\delta$-function is approximated by the Lorentzian distribution
with some small width. 
Since we need all the eigenvalues and eigenvectors, the cluster size
of the model was limited by the available disk space of the computer.
We show in the following the results for the $\sqrt {10} \times \sqrt
{10}$  cluster under the periodic  boundary condition.

\subsection{NMR relaxation rate}

In order to obtain the temperature dependence of the NMR relaxation
rate, we evaluate $1/T_1T$ by the following formula:
\begin{equation}
\label{t1t}
{1\over {T_1T}} = \lim_{\omega \to 0}{1\over {\omega}}\sum_q {\rm
Im}
\chi (q,\omega) .
\end{equation}
The effect of the $q$-dependence of the hyperfine form factor is
neglected, for simplicity. To clarify the role of antiferromagnetic
spin correlation, we have also evaluated the $T$ dependence of the
$Q\equiv (\pi, \pi)$ component of the spin correlation function,
\begin{equation}
S(Q)={1\over {N}}\sum _{{\bf i},{\bf j}}
(-1)^{({\bf j}-{\bf i})\cdot(\hat x +\hat y)}\langle S^z_{\bf i} S^z_{\bf j} 
\rangle .
\label{sq}
\end{equation}
In our previous study\cite{sakai} the following nearest-neighbor
spin correlation function was used instead for the purpose, 
\begin{equation}
\label{cor}
C_1={1\over {N}}\sum _{\bf i}{1\over 4}
\sum_{\rho=\pm \hat x,\pm \hat y}\langle 
S^z_{\bf i} S^z_{{\bf i}+\rho} \rangle .
\end{equation}
We are using $S(Q)$ here to emphasize the continuity that there exists
the antiferromagnetic order in the limit of no doping.  Note that the
terms, {\em magnetically singlet} and {\em antiferromagnetic}, are used
in nearly the same meaning. We are only concerned with the short range
magnetic correlation of the system of finite clusters. According to
the above definitions there is no reason to make definite
distinction between $S(Q)$ and $C_1$.  Actually both of $S(Q)$ and
$C_1$ show almost no essential difference in their temperature
dependence\cite{sakai2}. 

\begin{figure}
\begin{center}
\mbox{\psfig{figure=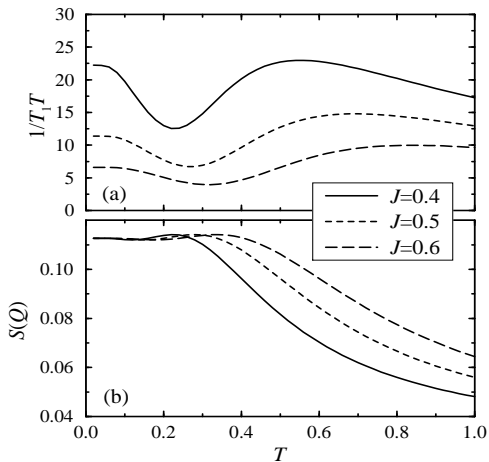,width=8cm,height=6cm,angle=0}}
\caption{
Temperature dependence of (a) $1/T_1T$ and (b) $S(Q)$ for $\delta =0$. 
}
\label{fig1}
\end{center}
\end{figure}

We show in Fig.\ref{fig1} results of $T$ dependence of (a) $1/T_1T$
and (b) $S(Q)$ for the undoped case ($\delta =0$) for $J$=0.6, 0.5,
and 0.4.  Only for $J$=0.4, the temperature dependence
of $S(Q)$ has already been reported by Tohyama {\it et al}.\cite{tohyama}
For cuprate superconductors the value $J=0.4$ is estimated
experimentally.  All the calculated $T$ dependence of $1/T_1T$
exhibits a broad peak around $T\sim J$ where $S(Q)$ shows the
significant increase. 
The above behavior of $1/T_1T$ for $\delta=0$ is, for instance, 
consistent with the high temperature series expansion for the
Heisenberg model.\cite{chakravarty,singh} 
The quantum Monte Carlo
simulation study by Sandvik and Scalapino\cite{sandvik}, however,
showed no peak behavior in the $1/T_1T$-$T$ curve. No peak is also
observed by the NMR measurement for the undoped
La$_2$CuO$_4$.\cite{imai} The reason is the temperature in these
studies is not so high enough to cover the crossover region of the
order of $J$.

\begin{figure}
\begin{center}
\mbox{\psfig{figure=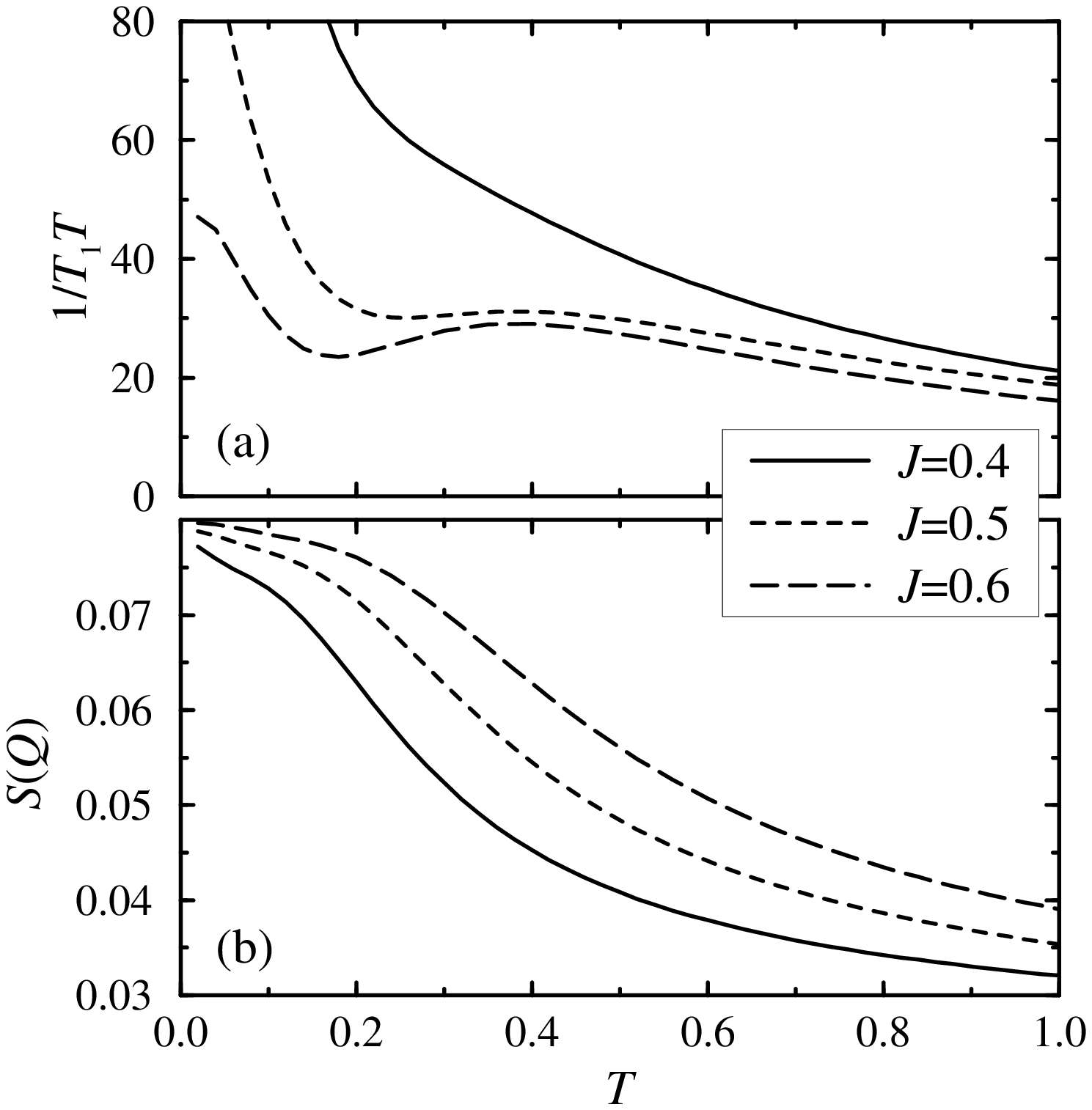,width=8cm,height=6cm,angle=0}}
\caption{
Temperature dependence of (a) $1/T_1T$ and (b) $S(Q)$ for $\delta =0.1$.
} 
\label{fig2}
\end{center}
\end{figure}

\begin{figure}
\begin{center}
\mbox{\psfig{figure=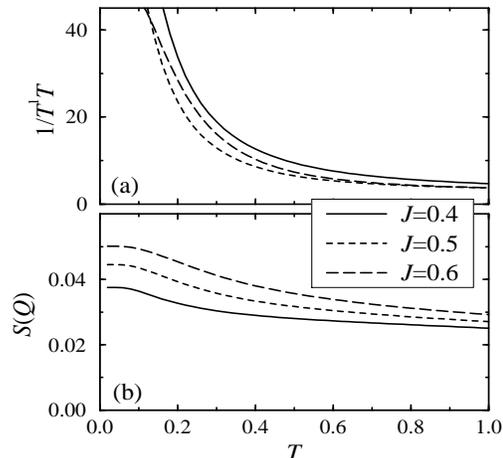,width=8cm,height=6cm,angle=0}}
\caption{ 
Temperature dependence of (a) $1/T_1T$ and (b) $S(Q)$ for $\delta =0.2$.
}
\label{fig3}
\end{center}
\end{figure}

The same results are shown in Figs.\ref{fig2}(a) and \ref{fig2}(b) for the
one-hole case ($\delta =0.1$). Peaks in the $T$ dependence of $1/T_1T$
for $J$=0.5 and 0.6 are shifted to lower temperature where the rapid
growth of $S(Q)$ is observed, in agreement with our mechanism.
Although no peak appears for the case $J$=0.4, a slight hump still
exists around the same temperature.  Such a behavior will be detected
as a deviation from the Curie-Weiss like temperature dependence of
$1/T_1T$, rather than a peak as was actually observed for
La$_{2-x}$Sr$_x$CuO$_4$.\cite{fujiyama} The same calculation for the
hole-doped 4$\times$4 $t$-$J$ cluster\cite{jaklic} does not show any
gap-like anomalies.  This is probably because the temperature range
didn't cover the region where the pseudogap is expected to appear.
The present results indicate that the gap-like behavior appears at
much higher temperature than the calculated lowest excitation gap
$\Delta $ of the cluster.  For example, the broad peak appears around
$T\sim 0.4$, while the estimated $\Delta$ is 0.146, for $J=0.6$ and
$\delta$=0.1.\cite{sakai} Therefore it excludes the possibility that
the pseudogap is caused by the discreteness of energy levels of finite
systems.
The same results for the two-hole system ($\delta =0.2$) are shown in
Figs.\ref{fig3}(a) and \ref{fig3}(b).  Although no gap-like behavior is
observed, they are also consistent with our view because of the
absence of any significant enhancement of the antiferromagnetic spin
correlation.

\subsection{Neutron scattering intensity}

\begin{figure}
\begin{center}
\mbox{\psfig{figure=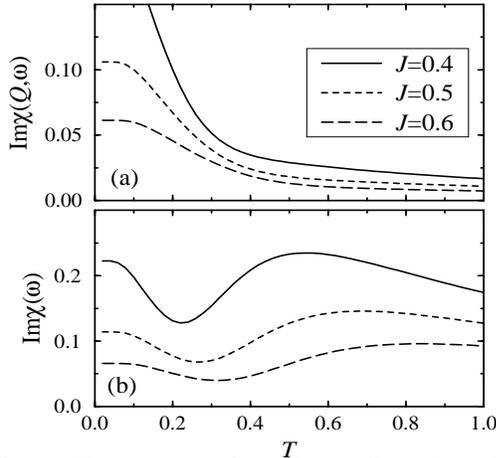,width=8cm,height=6cm,angle=0}}
\caption{
  Temperature dependence of (a) ${\rm Im}\chi (Q,\omega)$ and (b)
  ${\rm Im}\chi  (\omega)$ for $\delta =0$.  }
\label{fig4}
\end{center}
\end{figure}

\begin{figure}
\begin{center}
\mbox{\psfig{figure=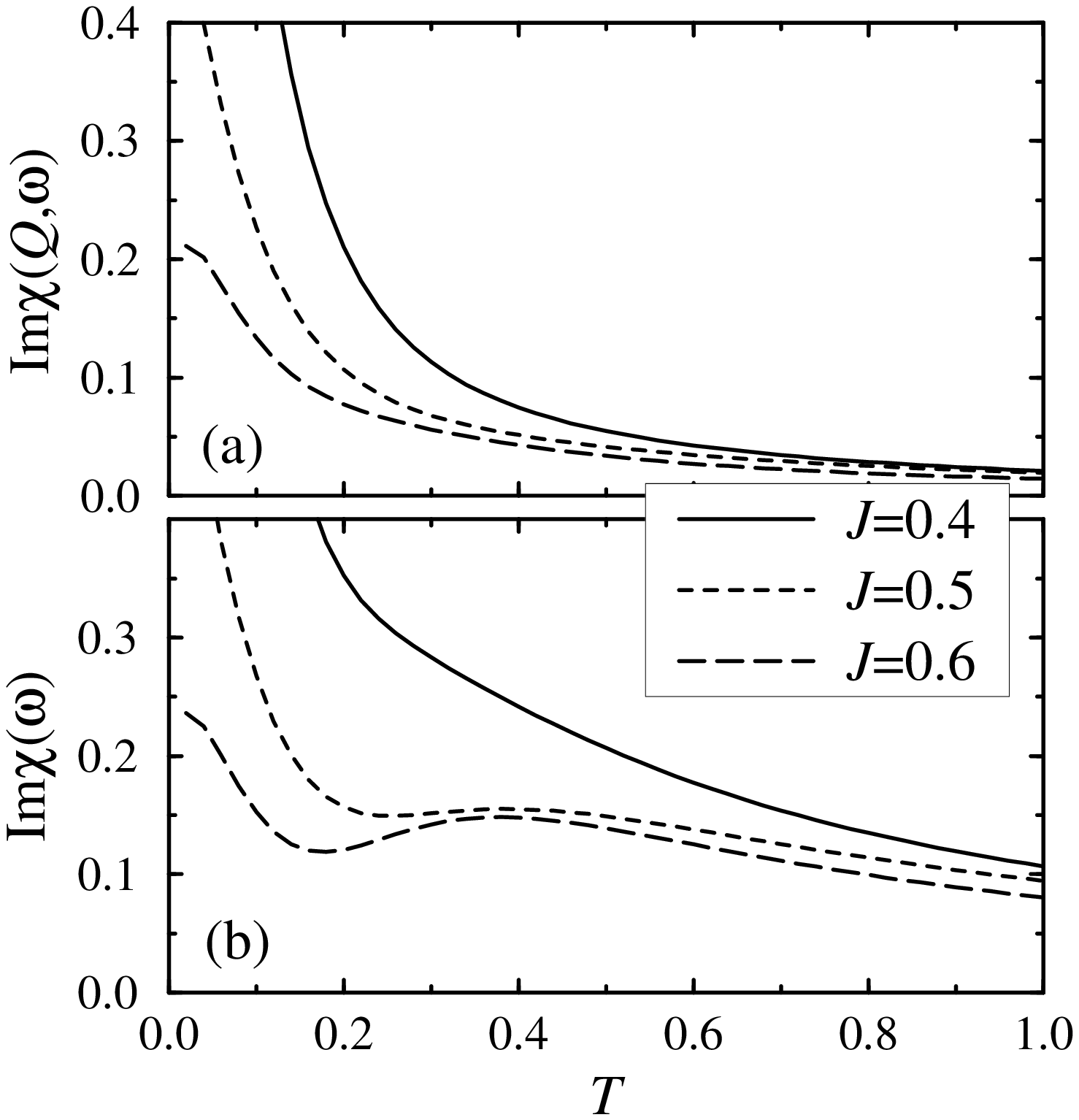,width=8cm,height=6cm,angle=0}}
\caption{
  Temperature dependence of (a) ${\rm Im}\chi (Q,\omega)$ and (b)
  ${\rm Im}\chi  (\omega)$ for $\delta =0.1$.  }
\label{fig5}
\end{center}
\end{figure}

\begin{figure}
\begin{center}
\mbox{\psfig{figure=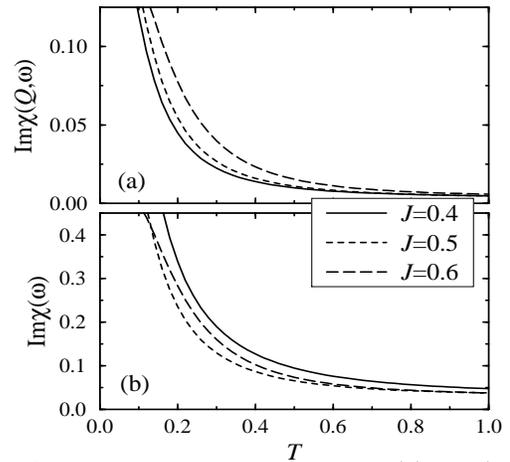,width=8cm,height=6cm,angle=0}}
\caption{
  Temperature dependence of (a) ${\rm Im}\chi (Q,\omega)$ and (b)
  ${\rm Im}\chi  (\omega)$ for $\delta =0.2$.  }
\label{fig6}
\end{center}
\end{figure}

The measurement of the neutron scattering intensity of
YBa$_2$Cu$_3$O$_{6.6}$ by Sternlieb {\it et al}. \cite{neutron2} indicated the
presence of the pseudogap behavior in the temperature dependence of the
$q$-integrated dynamical susceptibility
\begin{equation}
{\rm Im} \chi (\omega )\equiv 
\int {\rm d}q {\rm Im} \chi (q,\omega). 
\label{integrate}
\end{equation}
We show, in Figs.\ref{fig4}, \ref{fig5} and \ref{fig6}, the
temperature dependence of (a) ${\rm Im}\chi (Q,\omega)$ and (b) ${\rm
  Im}\chi(\omega)$ for a small $\omega$ (=0.01$t$) for $\delta$=0, 0.1
and 0.2.  Figs.\ref{fig4}(a), \ref{fig5}(a) and \ref{fig6}(a) show
the monotonic increase of ${\rm Im}\chi(Q,\omega)$ for all the values
of $J$ with decreasing $T$, independent of the doping $\delta$.  On
the other hand, $q$-integrated local susceptibility in
Figs.\ref{fig4}(b), \ref{fig5}(b) and \ref{fig6}(b) reveals a
gap-like behavior around the same crossover temperature as that of
$1/T_1T$ for $\delta \leq 0.1$. These results are in good agreement
with the above neutron scattering experiment. As for the former, the
low-frequency limit of ${\rm Im}\chi(Q,\omega)$ is given by
\[
{\rm Im}\chi(Q,\omega) \to \frac{\chi(Q,0)}{\Gamma_Q} 
 \propto \chi^2(Q,0), \quad (\omega \to 0), 
\]
where $\Gamma_Q$ is the Lorentzian distribution width of the frequency
spectrum.  The monotonic increase of ${\rm Im}\chi(Q,\omega)$ with
decreasing the temperature is therefore consistent with the absence of
the pseudogap behavior in the $T$-dependence of $1/T_{\rm 2G}$, i.e. 
$T_{\rm 2G}^2 \propto \chi^{-1}(Q,0)$, by the
NMR measurement.\cite{nmr2}

\subsection{Magnetic susceptibility}

\begin{figure}
\begin{center}
\mbox{\psfig{figure=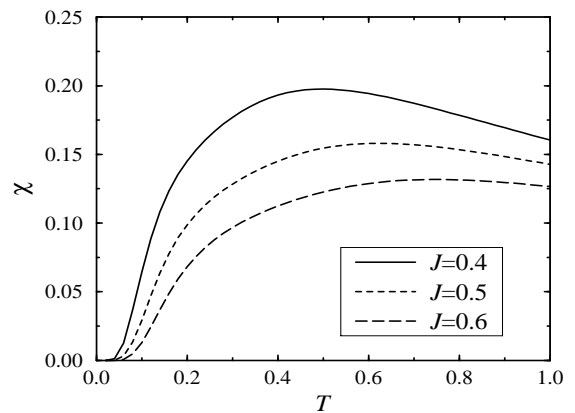,width=8cm,height=6cm,angle=0}}
\caption{
  Temperature dependence of $\chi$ for $\delta =0$.  }
\label{fig7}
\end{center}
\end{figure}

\begin{figure}
\begin{center}
\mbox{\psfig{figure=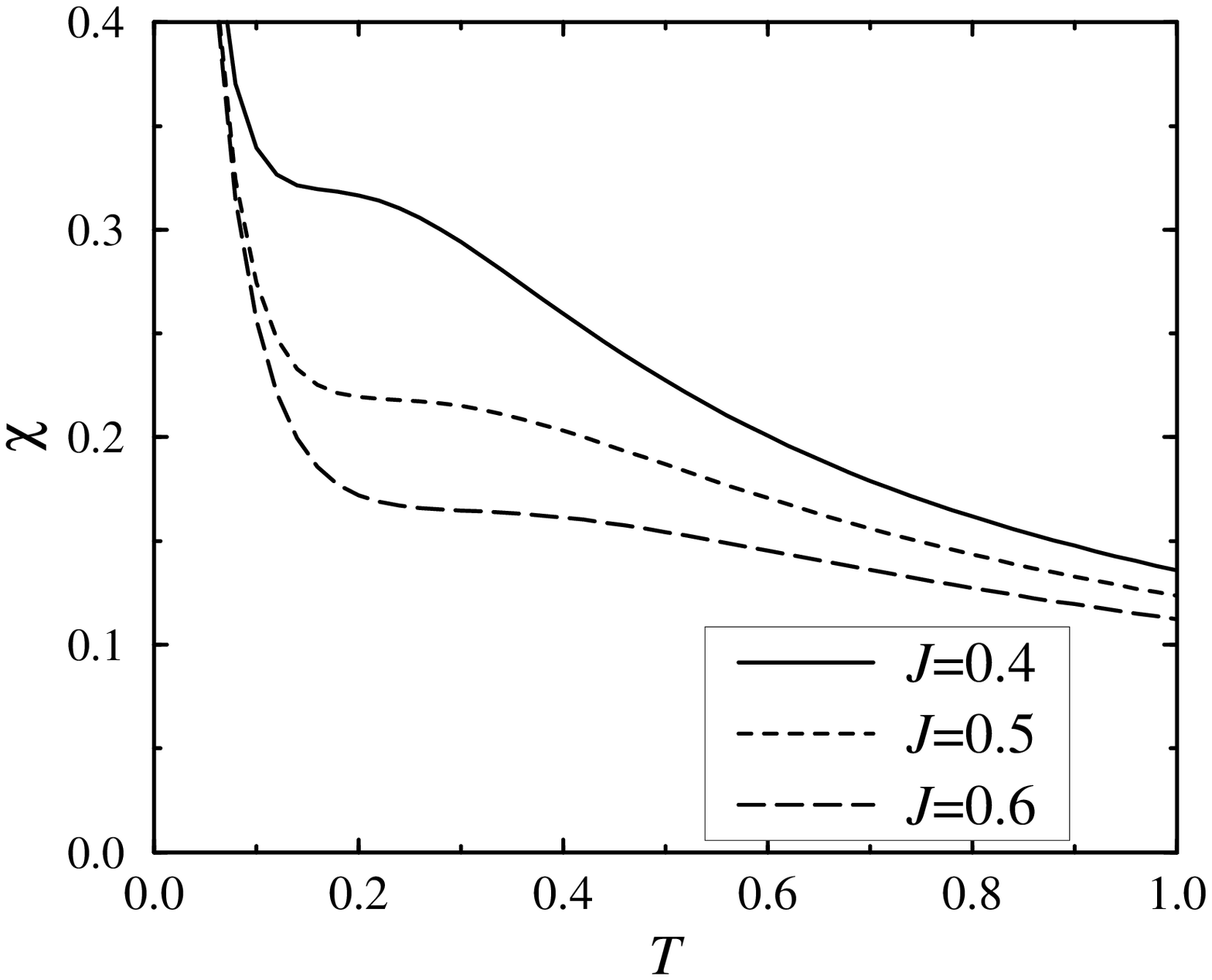,width=8cm,height=6cm,angle=0}}
\caption{
  Temperature dependence of $\chi$ for $\delta =0.1$.  }
\label{fig8}
\end{center}
\end{figure}

\begin{figure}
\begin{center}
\mbox{\psfig{figure=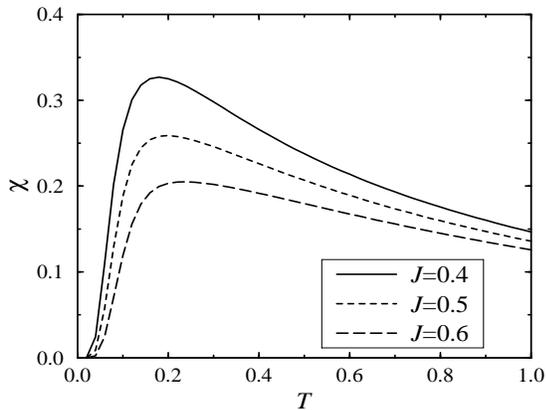,width=8cm,height=6cm,angle=0}}
\caption{
  Temperature dependence of $\chi$ for $\delta =0.2$.  }
\label{fig9}
\end{center}
\end{figure}

The pseudogap behavior has also been observed in the temperature dependence
of the static magnetic susceptibility $\chi$.\cite{chi} The calculated
results of $\chi$ for $\delta$=0, 0.1 and 0.2 are shown in
Figs.~\ref{fig7}, \ref{fig8} and \ref{fig9}, respectively. In all the
figures we can see the crossover behaviors around the same temperature
as those of $1/T_1T$ and ${\rm Im}\chi (\omega )$ in the underdoped
case ($\delta \leq 0.1$), even for the realistic value $J=0.4$.  As
shown in Fig.~9, peaks appear at lower temperature for $\delta=0.2$.
However they result from the different origin. If there are two holes
in a finite cluster model, its ground state always becomes
singlet and the total energy has some small excitation gap. Therefore
the susceptibility shows a peak around the temperature
corresponding to the lowest excitation gap of the system that comes
from these finite size effects. The behavior has nothing to do with
our mechanism. (The same calculation of $\chi$ for $J=0.4$ by Tohyama
{\it et al}.\cite{tohyama} yielded a peak even for $\delta =0.3$.)

To conclude, all the results presented above support our view that the
pseudogap behavior observed for various magnetic properties are induced
by the growth of the antiferromagnetic spin correlation.

\section{Next-Nearest Neighbor Hopping $t'$}

\begin{figure}
\begin{center}
\mbox{\psfig{figure=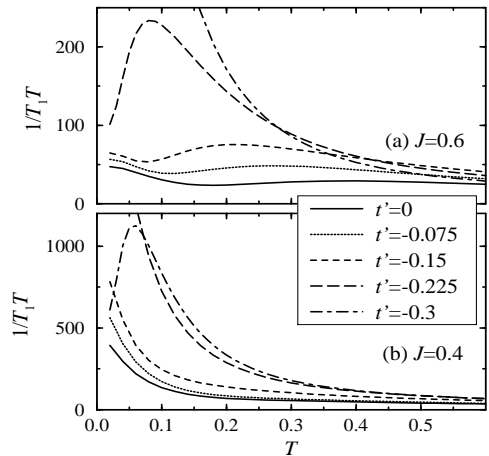,width=8cm,height=6cm,angle=0}}
\caption{
Temperature dependence 
of $1/T_1T$ for $\delta =0.1$ in the case of (a) $J=0.6$ and 
(b) $J=0.4$. 
}
\label{fig10}
\end{center}
\end{figure}

In the previous section we showed that the antiferromagnetic spin
correlation produces a gap-like behavior in the $T$-dependence of
$1/T_1T$, ${\rm Im}\chi(\omega)$ and $\chi$ around a common crossover
temperature.  The characteristic temperature for $\delta=0.1$ is
roughly estimated as $\sim J/2$, as shown in Figs. \ref{fig2}(a),
\ref{fig5}(b) and \ref{fig8}.  It is, however, much higher than
observed ones for real cuprates. This is partly because the effect of
the hole hopping $t$ is overestimated compared with that of $J$ due to
the finite size effect. As another possible reason, let us discuss
below the effect of including the next-nearest neighbor hole hopping,
\begin{eqnarray}
\label{next}
H' = - t' \sum_{<\bf i,\bf j>', \sigma}
            ({c}_{\bf j,\sigma}^\dagger {c}_{\bf i,\sigma}
           + {c}_{\bf i,\sigma}^\dagger {c}_{\bf j,\sigma} ), 
\end{eqnarray}
into the $t$-$J$ Hamiltonian (\ref{ham}), where $\sum_{<\bf i,\bf
  j>'}$ means the summation over all the next-nearest sites.  The term is
originally introduced to explain the dynamics of the correlated motion
of spin and hole-hopping degrees of freedom in some cuprates. The
value $t' \sim -0.3$ is estimated to be suitable for
Sr$_2$CuO$_2$Cl$_2$ by the exact diagonalization study.\cite{next} To
clarify the effect of $t'$ on the pseudogap formation, we show the $T$
dependence of $1/T_1T$ for various values of $t'$ in
Figs.~\ref{fig10}(a) for $J=0.6$ and \ref{fig10}(b) 
for $J=0.4$ ($\delta=0.1$).
From Fig.~\ref{fig10}(a) we can see that the peak temperatures are
lowered by the effect. The narrowing of the peak width takes place at
the same time. As the result the slight anomaly in Fig.\ref{fig2}(a)
for $J=0.4$, shifted to lower temperature, now becomes evident if we
assume $t'=-0.3$ as shown Fig.\ref{fig10}(b). Therefore it is possible
to predict the crossover temperature around a realistic temperature
range $\sim J/4$ by including the effect of the next-nearest neighbor
hopping.

As we increases the value of $t'$, there appears, however, another
difficulty. Because the peak temperature is lowered and becomes
comparable with the order of the lowest excitation gap of the cluster,
it is very difficult to distinguish the peak behavior from the one
that comes from the discreteness of the energy levels of the system.
Further calculations on larger size clusters will be necessary to
clarify the situation and to obtain reliable conclusions.

\section{Discussion}

The results of the present study clearly showed the presence
of gap-like behaviors in the temperature dependence of various
magnetic properties around a single crossover temperature due to the
development of antiferromagnetic spin correlation.  Experimentally
detailed gap-like behaviors differ slightly with each other depending
on materials. In the case of La$_{2-x}$Sr$_x$CuO$_4$, for instance,
the crossover temperature of $\chi$ ($T^*_{\chi}$)\cite{chi} is
slightly higher than that of $1/T_1T$ ($T^*_{\rm NMR}$)\cite{nmr3}.
The recent NMR Knight shift measurement for
Bi$_2$Sr$_2$CaCu$_2$O$_{8+\delta}$ also indicates the presence of two
different crossover temperatures.\cite{nmr4} Stimulated by these
discoveries, it is argued that the higher crossover temperature (the
large pseudogap) will be associated with the growth of the
antiferromagnetic spin correlation.  On the other hand, the lower one
(the small pseudogap) will be due to the singlet (or superconducting
pair) formation, i.e. $T^*_{\chi}$ will be assigned to the former,
while $T^*_{\rm NMR}$ the latter. From our approach
based on the finite cluster model, it will be difficult to predict the
presence of double pseudogap phenomena.

Our model Hamiltonian employed here is quite simplified one. In order
to deal with actual systems we have to include various additional
interactions specific to each system. In the present paper we are
particularly interested in properties that are common to all the
cuprate superconductors. Therefore in order to compare with
experiments, we have to be aware that the observed properties are
intrinsic to all of them or not. It will also be important to confirm
whether the presence of two crossover temperatures is common to all
the cuprates systems.

As was pointed out in the Introduction, spin-gap phenomena is
characteristic to the underdoped cuprates where the superconducting
transition temperature $T_{\rm c}$ decreases
proportional to the hole doping concentration. On the other hand the
pseudogap temperature increases.  Because of this opposite doping
dependence, it seems to be difficult to associate the pseudo-gap
phenomena with the superconducting pairing fluctuations.
The antiferromagnetic correlation, therefore, plays
the dominant role on reducing the low energy excitations of the system
as far as the magnetic freedoms are concerned.  Only around the
critical temperature $T_{\rm c}$, they will be further modified by the
presence of the superconducting transition. These effect are limited
to the very low energy regions.  The present mechanism is based on the
well-known properties of low-dimensional Heisenberg magnets.  The
overall doping dependence of the crossover temperature is also in
agreements with experiments. It decreases with increasing hole doping,
whereas the transition temperature $T_{\rm c}$ increases towards the optimum
condition.  Because of these reasons we suppose that our mechanism is
one of most probable candidates for the phenomena.
 
Because the model treated here is a single-hole system ($\delta=0.1$
for the 10 site cluster), its ground state is doublet. Even in this
doublet case we can derive the gap-like behavior in the temperature
dependence of $1/T_1T$. If the same calculation is done on a two-hole
system of a larger cluster with the same doping $\delta$, a slightly
different properties will be obtained because its ground state now
becomes singlet.  We expect that the gap-like behavior will then be
more evident.  The difference becomes less conspicuous, if we increase
the size of clusters, for it results from the size effect.  Though it
is very difficult because of various limitations concerning the 
available resources of computer systems, it will be interesting to
extend our present numerical diagonalization study to the two-hole
system with larger cluster size to clarified the situation.

\section{Summary}
We have shown that the pseudogap behaviors observed in the
high-temperature superconductors originate from the growth of the
antiferromagnetic spin correlation based on the exact diagonalization
study  on the finite cluster $t$-$J$ model in a realistic parameter
region for the real cuprates.   We also found  that the pseudogap
temperature is lowered by including the next-nearest neighbor hopping
effect.

\section*{Acknowledgment}
We wish to thank Profs M. Sato, K Yamada,  Y. Ohta and Y. Hasegawa for
fruitful discussions.  We also thank the Supercomputer Center,
Institute for Solid State Physics, University of Tokyo for the
facilities and the use of the Fujitsu VPP500.  This research was
supported in part by Grant-in-Aid for the Scientific Research Fund
from the Ministry of Education, Science, Sports and Culture (11440103).

\end{document}